\documentstyle[twoside,fleqn,npb,epsfig]{article}
\newcommand{\AmS}{{\protect\the\textfont2
  A\kern-.1667em\lower.5ex\hbox{M}\kern-.125emS}}

\title{Full $O(\alpha)$ corrections to $e^+e^-\rightarrow $
       $\nu \bar{\nu} H$ by {\tt GRACE}}

\author{G. B\'{e}langer\address{LAPTH, B.P.110, Annecy-le-Vieux F-74941,
                                France},
        F. Boudjema$^a$,
        J. Fujimoto\address{KEK, Oho 1-1, Tsukuba, Ibaraki 305-0801, Japan},
        T. Ishikawa$^b$,
        T. Kaneko$^b$,
        K. Kato\address{Kogakuin University, Nishi-Shinjuku 1-24,
        Shinjuku, Tokyo 163-8677, Japan}
        and
        Y.Shimizu$^b$}

\newcommand{\nnht}{$e^+ e^- \rightarrow \nu \bar{\nu} H \;$}
\begin{document}

\begin{abstract}
We present the full $O(\alpha)$ corrections to single Higgs
production in $e^+e^-$ collision. The computation is performed
with the help of {\tt GRACE-LOOP} where a generalized non-linear
gauge fixing condition is implemented. The numerical results are
checked by testing their UV and IR finiteness as well as their
independence on all five non-linear gauge parameters. We find that
for a 500 GeV collider and a light Higgs boson of mass 150GeV, the
total correction is small when the results are expressed in terms
of $\alpha$ rather than $G_\mu$. For a higher Higgs boson mass of
350GeV, the correction is of order $-10$\%.

\end{abstract}

\maketitle
\vspace*{-11cm}
\noindent
KEK-CP-134 \\
LAPTH-Conf-947/2002
\vspace*{8.7cm}

\section{Introduction}

Single Higgs production  will be one of the most important
processes at the future electron-positron linear collider. There
are two main mechanisms to this reaction: Higgs-strahlung and $W$
fusion. At tree-level, the calculation is well under control and
takes into account the interference effects between Higgs-strahlung
and $W$ fusion\cite{zerwas}. The one-loop corrections to the
two-body $e^+e^- \rightarrow ZH$ Higgs-strahlung process have been
studied both in the standard model\cite{denner} and  in the
minimal supersymmetric standard model\cite{hollik}. However, despite 
the fact that for linear collider energies  $W$ fusion becomes the
dominant Higgs production mechanism a full $O(\alpha)$ calculation 
has been missing.  This is indispensable if one wants to carry a
precision measurement program. The loop correction to the $HWW$
vertex has been considered \cite{kniehl} and one could argue that
it might constitute a good approximation for the fusion process,
but it rests to see how well this approximation fares in
comparison with the full calculation.  Very recently one-loop
radiative corrections to the fusion process have been investigated
within supersymmetry\cite{vienna,heinemeyer} but again by only
taking into account the contribution of the fermions and sfermions
to the $H/hWW$ vertex. In this paper, we report the full $O(\alpha)$
corrections consisting of virtual and soft corrections as well as
hard photon radiation, to single Higgs production in $e^+e^-$,
including both the fusion and Higgs-strahlung processes in the
standard model.\footnote{In the workshop, O. Tarasov also
presented the same topics.}

\section{{\tt GRACE-LOOP} system}

The Feynman
diagrams and the corresponding matrix elements 
for \nnht
were generated by
{\tt GRACE-LOOP}\cite{grace-loop}, an automatic calculation system.
In addition, the radiative process 
$e^+e^- \rightarrow \nu\bar{\nu} H \gamma$ 
with a hard photon has been calculated by
{\tt GRACE}\cite{grace}. The latter process is also necessary to
complete the total correction.
We deal with two sets of diagrams, the {\itshape full set\/} 
and the {\itshape production set\/}.
In the standard model and within the non-linear gauge fixing
conditions which will be discussed later, 
one counts 12 tree-level and 1350 1-loop diagrams (with 98 pentagon
graphs). This defines the {\itshape full set\/}. When the diagrams including
scalar/pseudoscalar-electron couplings proportional to the electron
mass are removed
from the {\itshape full set\/}, 
we obtain the {\itshape production set\/} which consists of
2 tree diagrams and 249 1-loop diagrams (with 15 pentagon
graphs).  
The {\itshape full set\/} has been used to make careful checks of the
numerical calculation in quadruple precision.
After checking the system with the {\itshape full set\/} at some random
points in phase space,  
the phase space integration of the {\itshape production set\/}
is done in double precision 
to obtain cross sections and distributions. 
The {\itshape production set\/} is sometimes used in quadruple precision
to confirm the numerical stability of the result.

The {\tt GRACE-LOOP} system, which has
a model file that describes all the interaction vertices derived
from a particular Lagrangian, can generate all the necessary
Feynman graphs together with their codes, so that matrix elements
can be generated for the calculation of the
cross section and  event generation. 
For loop processes, the
automatic calculation goes through an intermediate step performing
some symbolic manipulation. This involves {\tt
REDUCE}\cite{reduce} or {\tt FORM}\cite{form} for  handling all
the Dirac and tensor algebra in $n$-dimension for all the
interference terms between tree-level and 1-loop diagrams. 
The one-loop diagram contribution from each loop graph $g$
at a phase space point, is defined as
\begin{equation}
{\rm d}\sigma_g \propto 2\Re \left(T^{loop}_g\cdot
T^{tree\ \dagger}\right)
\label{eq:sigma}
\end{equation}
where $T^{tree}$ is the tree-level amplitude summed over all
tree-diagrams. $T^{loop}_g$ is the one-loop amplitude
contribution of a  one-loop diagram $g$. 
The Feynman trick for the propagator is also automatically applied at
this stage. This is then passed to a module of
libraries for the loop integration containing the {\tt FF}\cite{ff} 
packages as well as the in-house reduction formulas. The
system together with the one-loop renormalization set-up has been
described in \cite{nlg}.

\section{Checks of the system}

The set of input parameters for the calculation is the following. 
Throughout this paper, the results are expressed in terms of the fine
structure constant in the Thomson limit $\alpha_{QED}=137.0359895$
and the $Z$ mass $M_Z=91.1876$GeV. The on-shell renormalization
program uses $M_W$ as input parameter, nonetheless the numerical
value of $M_W$ is derived through $\Delta r$\cite{Hioki}
\footnote{We include NLO QCD corrections and two-loop Higgs
effects. We take $\alpha_s(M_Z^2)$=0.118 together with $G_{\mu}$
=1.16639$\times 10^{-5} GeV^{-2}$.}. $M_W$ thus changes as a
function of $M_H$. We take the set $M_u$=$M_d$=58MeV, $M_s$=92MeV,
$M_c$=1.5GeV, $M_b$=4.7GeV, which gives a ``perturbative" value of
$\alpha(M_Z)$ compatible with the current experimentally driven
value. We also take $M_{top}$=174GeV. With theses values, we
derive $M_W$ =80.3767GeV for $M_H$=150GeV, and $M_W$=80.3158GeV
for $M_H$=350GeV. For the Higgs-strahlung subprocess we require a
$Z$-width. We have taken a constant $Z$-width,
$\Gamma_Z$=2.4952GeV. Unless  otherwise stated our results refer
to the full $e^-e^+\rightarrow \nu \bar{\nu} H$, summing over all
three types of neutrinos with, for electron neutrinos, the effect
of interference between fusion and Higgs-strahlung.

The results of the calculation are checked by performing four
kinds of tests at some point in phase space. For these tests to be
passed one works with the {\itshape full set\/}
in quadruple precision. One first checks the
ultraviolet finiteness of the results. The regulator constant
$C_{UV}=1/\varepsilon -\gamma_E+\log 4\pi, n=4-2\varepsilon$ is
kept in the matrix elements. When one varies this parameter
$C_{UV}$, the ultraviolet finiteness test gives a result that is
stable over 30 digits. Infrared (IR) finiteness is checked by
introducing a  fictitious photon mass $\lambda$, treated in the
code as an input parameter. The sum of loop and bremsstrahlung
contributions is stable over 23 digits when varying $\lambda$. The
third check relates to the independence on the parameter $k_c$
which is a soft photon cut parameter that separates soft photon
radiation (analytical formula) and the hard photon performed  by
the Monte-Carlo integration. Gauge parameter independence of the
result is performed as a last check through a set of five gauge
fixing parameters. For the latter a generalized non-linear gauge
fixing condition\cite{boudjema} has been chosen.

\begin{eqnarray}
{\cal L}_{\rm GF}=\hspace*{51mm} \nonumber \\
-{1\over\xi_W}\biggl|(\partial_{\mu}-ie\tilde{\alpha}A_{\mu}
-ig\cos\theta_W\tilde{\beta}Z_{\mu})W^{+\mu}  \nonumber \\
 \qquad \qquad 
+\xi_W{g\over2}(v+\tilde{\delta}H+i\tilde{\kappa}\chi_3)\chi^{+}
\biggr|^2  \nonumber \\
-\frac{1}{2\xi_Z}\left(\partial_{\mu}Z^{\mu}
+\xi_Z\frac{g}{2\cos\theta_W}(v+\tilde{\varepsilon}H)\chi_3\right)^2 
  \nonumber \\
-{1\over2\xi_A}(\partial_\mu A^\mu)^2
\end{eqnarray}

\begin{table*}[hbt]
\caption{Numerical size of S/M$^{(i)}$ for each non-linear gauge
parameter. '\# of graphs' means the number of graphs depending
on $\tilde{\zeta}$. Except for $\tilde{\delta}$, no 3rd order
term appears in this process.
}

\begin{center}
\begin{tabular}{|c|c|c|c|c|}
\hline
\rule[-1mm]{0mm}{5mm} $\tilde{\zeta}$ 
& $\#$ of graphs & ${\rm S/M}^{(3)}$ & ${\rm S/M}^{(2)}$ & ${\rm S/M}^{(1)}$ \\
\hline
\rule[-1mm]{0mm}{5mm} $\tilde{\alpha}$ &
149  &  $-$ & $10^{-28}$ & $10^{-30}$  \\
\hline
\rule[-1mm]{0mm}{5mm} $\tilde{\beta}$ &
314  &  $-$ & $10^{-31}$ & $10^{-23}$  \\
\hline
\rule[-1mm]{0mm}{5mm} $\tilde{\delta}$ &
1059  & $10^{-20}$  & $10^{-20}$ & $10^{-26}$  \\
\hline
\rule[-1mm]{0mm}{5mm} $\tilde{\kappa}$ &
122  & $-$ & $10^{-23}$ & $10^{-23}$  \\
\hline
\rule[-1mm]{0mm}{5mm} $\tilde{\varepsilon}$ &
132  & $-$ & $10^{-21}$ & $10^{-30}$  \\
\hline
\end{tabular}
\end{center}
\end{table*}

The $\chi$ represent the Goldstone. We take 
$\xi_W=\xi_Z=\xi_A=1$ so that no ``longitudinal" term
appears in the gauge propagators. Not only this makes the
expressions much simpler and avoids unnecessary large
cancellations, but it also avoids the need for high tensor
structures in the loop integrals. The use of five parameters is
not redundant as often these parameters check complementary sets
of diagrams. 
We introduce a generic notation $\tilde{\zeta}$ to represent either of
$\tilde{\alpha}, \tilde{\beta}, \tilde{\delta}, \tilde{\kappa}, \tilde{\epsilon}$.
For each $\tilde{\zeta}$ the first check is made while freezing all 
other four parameters to be $0$. 
We have also made checks with two parameters non-zero. 
In principle checking for $2$ or $3$ values of the gauge parameter
should be convincing enough. We in fact go one step further and
perform a complete test of gauge parameter independence. To
achieve this we generate for each non-linear gauge parameter
$\tilde{\zeta}$, the values of the loop correction to the total
differential cross section as well as the contribution of each
one-loop diagram contribution for the five values $\tilde{\zeta}=0,\pm 1,
\pm 2$.
A rapid look at the
structure of the Feynman rules of the non-linear gauge leads one
to conclude that for \nnht each contribution is a polynomial of
(at most) third degree in the gauge parameter and thus, that each
contribution, ${\rm d}\sigma_g$ in Eq.(\ref{eq:sigma})
may be written as

\begin{equation}
{\rm d}\sigma_g(\tilde{\zeta})={\rm d}\sigma_g^{(0)}
+\tilde{\zeta} {\rm d}\sigma_g^{(1)}
+\tilde{\zeta}^2 {\rm d}\sigma_g^{(2)} +\tilde{\zeta}^3 {\rm d}\sigma_g^{(3)}
\end{equation}

For each contribution ${\rm d}\sigma_g$, it is a straightforward
matter, given the values of ${\rm d}\sigma_g$ for the five input
$\tilde{\zeta}=0,\pm 1, \pm 2$, to reconstruct $\sigma_g^{(0,1,2,3)}$.
For each set of parameters we
automatically pick up all those diagrams that involve an explicit
dependence on the gauge parameter. 
The number
of diagrams in this set depends on the parameter chosen. In some
cases a huge number of diagrams is involved. For the process at
hand this occurs with the parameter $\tilde{\delta}$ where over
$1000$ diagrams contribute to the sums, as shown in Table 1, and thus almost 
the entire set of diagrams, is involved in the check.

We then verify that the differential cross section is independent
of $\tilde{\zeta}$
\begin{equation}
{\rm d}\sigma=\sum_g {\rm d}\sigma_g(\zeta)=\sum_g {\rm d}\sigma_g^{(0)}.
\end{equation}

\noindent
In order to check this, we introduce the following notation for each parameter,

\begin{equation}
{\rm S/M}^{(i)}=\frac{\sum_g{\rm d}\sigma_g^{(i)}}{{\rm Max}_g\bigl( |{\rm
d}\sigma_g^{(i)}|\bigr)}\;\; ,\;\; i=1,2,3.
\end{equation}

As seen from Table 1 agreement within 20 to 30 digits is observed.
(The agreement gets better if one artificially gives the electron 
mass a higher value.)
The gauge parameter dependence check not only tests the
various components of the input file (correct Feynman diagrams for
example) but also the symbolic manipulation part and most
important of all the correctness of all the reduction formulae and
the proper implementation of all the $n$-point functions. This is
quite useful when one deals with 5-point functions as the case at
hand.

\section{Numerical results}

The results we show here is based on the {\itshape production set\/}
with all 3 neutrinos species and
include the hard bremsstrahlung part. Integration over all photon
energies and angles is thus performed. The effect of radiative
corrections are presented in the $\alpha$-scheme.

\begin{table*}[hbt]
\setlength{\tabcolsep}{1.5pc}
\newlength{\digitwidth} \settowidth{\digitwidth}{\rm 0}
\catcode`?=\active \def?{\kern\digitwidth}
\caption{Higgs mass dependence for the tree total cross sections, the full
$O(\alpha)$ corrected one and the radiative correction factor at $\sqrt{s}
=500$ GeV.}
\label{tab:effluents} \begin{tabular*}{\textwidth}{@{}l@{\extracolsep{\fill}}rrrr}
\hline
\cline{2-3} \cline{4-5}
\hline
$M_H$  & $\sigma_{tree}$ & $\sigma_{O(\alpha)}$ & $\Delta$  \\
(GeV)  & (fb) & (fb)        &     (\%)         \\
\hline
150 & 61.12& 60.99 $\pm$0.07 & $-0.2$  \\
200 & 37.33& 37.16 $\pm$0.04 & $-0.4$  \\
250 & 21.17& 20.63 $\pm$0.02 &  $-2.5$  \\
300 & 10.76& 10.30 $\pm$0.01 & $-4.2$  \\
350 & 4.603&4.184 $\pm$ 0.004  &$-9.1$  \\
\hline
\hline
\end{tabular*}
\end{table*}

\begin{figure}
\begin{tabular}{c}
\resizebox{!}{6.8cm}{\includegraphics*{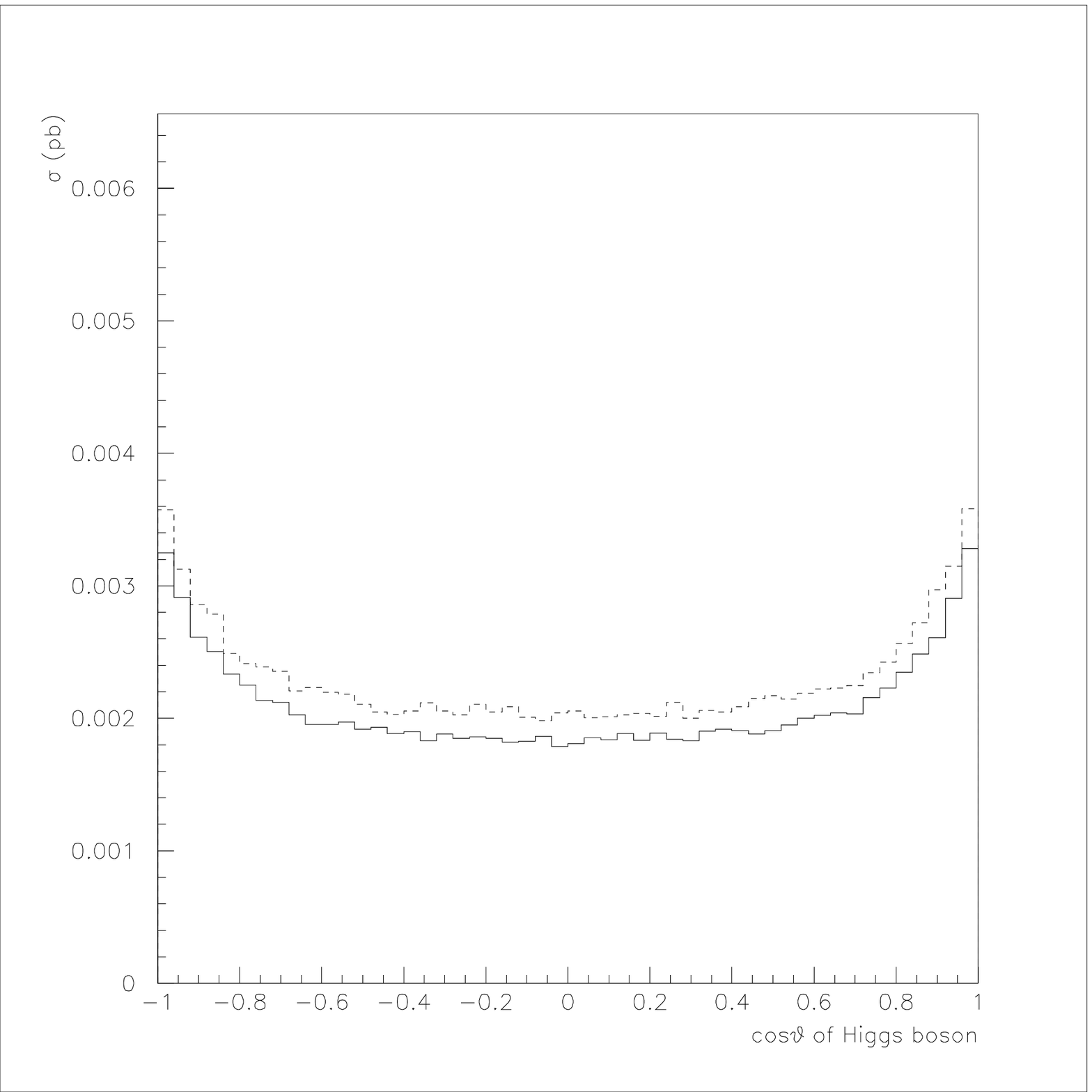}}
\end{tabular}
\caption{%
$\cos\theta$ distributions of Higgs boson at $\sqrt{s}=500$GeV
and $M_{Higgs}=350$ GeV. The solid and dashed lines show
$O(\alpha)$ corrected and tree level distributions, respectively. 
}
\end{figure}

\begin{figure}
\begin{tabular}{c}
\resizebox{!}{6.8cm}{\includegraphics*{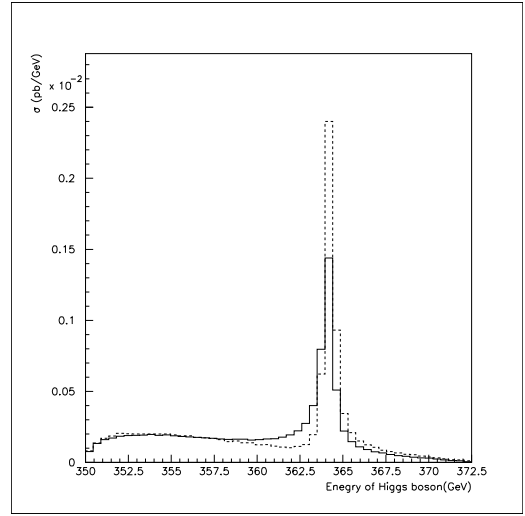}}
\end{tabular}
\caption{%
Energy distributions of Higgs boson at $\sqrt{s}=500$GeV
and $M_{Higgs}=350$ GeV. The solid and dashed lines show
$O(\alpha)$ corrected and tree level distributions, respectively. 
}
\end{figure}

The overall correction to the total cross section at the linear collider
energy ($\sqrt{s}$=500GeV), is small for a light Higgs mass of order
150 GeV as shown in Table 2. The radiative correction factor 
$\Delta$ is defined as
\begin{equation}
\Delta \equiv \frac{\sigma_{O(\alpha)}}{\sigma_{tree}} - 1.
\end{equation}

The magnitude of the correction increases steadily for higher Higgs masses. 
For example, for a Higgs mass of 350 GeV at a center-of-mass energy of
500 GeV the correction reaches $-10$\%. This $-10$\% correction is
distributed almost evenly over the range of  scattering angles of
the  Higgs boson, see  Fig.1.

The distribution of the Higgs boson energy is, however, affected
by the $O(\alpha)$ corrections only in the region of the resonance
of the $Z-$boson. The resonance structure of the tree level is
distorted in  the form of a radiative tail as shown in Fig. 2.

\section{Summary}

{\tt GRACE/LOOP}  has been applied to calculate the full
$O(\alpha)$ corrections to single Higgs production  in the
standard model. One-loop amplitudes of the full 1350 diagrams were
generated by this system. The non-linear gauge fixing condition
has been introduced and shown to be a very powerful tool to check
the consistency of the full set of amplitudes and the system
itself. At $\sqrt{s}$=500 GeV, the correction is around $-10$\%
for a 350 GeV Higgs mass. The radiative tail was observed for the
energy distribution of the Higgs boson. It means that for more
realistic predictions, one has somehow to include multiple photon
emissions.

\section{Acknowledgment}

This work is part of a collaboration between the {\tt GRACE}
project in the Minami-Tateya group and  LAPTH. D. Perret-Gallix
and Y. Kurihara deserve special thanks for their contribution.
They also would like to acknowledge the local organizing committee
of RADCOR 2002/LOOPS and LEGS 2002 for a stimulating workshop and
for  their nice organization. This work was supported in part by
Japan Society for Promotion of Science under the Grant-in-Aid for
scientific Research B(no. 1440083) and PICS 397 of the French
National Centre for Scientific Research.


\end{document}